\documentstyle[manuscript,eqsecnum,aps,psfig,cite]{revtex}

\begin{document}
\draft
\tighten
\title{Efficiency of different numerical methods for solving 
Redfield equations}
\author{Ivan Kondov, Ulrich Kleinekath\"ofer, and Michael Schreiber}
\address{Institut f\"ur Physik, Technische Universit\"at,
D-09107\ Chemnitz, Germany}
\maketitle
\begin{abstract}
  The numerical efficiency of different schemes for solving the
  Liouville-von Neumann equation within multilevel Redfield theory has been
  studied.  Among the tested algorithms are the well-known Runge-Kutta
  scheme in two different implementations as well as methods especially
  developed for time propagation: the Short Iterative Arnoldi, Chebyshev
  and Newtonian propagators. In addition, an implementation of a symplectic
  integrator has been studied.  For a simple example of a two-center
  electron transfer system we discuss some aspects of the efficiency of
  these methods to integrate the equations of motion. Overall for
  time-independent potentials the Newtonian method is recommended. For
  time-dependent potentials implementations of the Runge-Kutta algorithm
  are very efficient.

\end{abstract}


\section{Introduction}
\label{sec:intro}
Besides classical and semi-classical descriptions of dissipative molecular
systems several quantum theories exist which fully account for the quantum
effects in dissipative dynamics. Among the latter are the reduced density
matrix (RDM) formalism \cite{blum96,may00,davi98b,kosl97,kohe97a} and the
path integral methods \cite{weis99,makr98}. Here we concentrate on the
Redfield theory \cite{redf57,redf65}, in which one has to solve a master
equation for the RDM. It is obtained by performing a second order
perturbation treatment in the system-bath coupling as well as restricting
the calculation to the Markovian limit.  With this approach the quantum
dynamics of an ``open'' system, e.g., the exchange of energy and phases
with the surroundings modeled as a heat bath, can be described. The
unidirectional energy flow into the environment is called dissipation.
Within this theory it is possible e.g.\ to simulate the dissipative
short-time population dynamics usually detected by modern ultrafast
spectroscopies.  

Because the Redfield theory is a Markovian theory the time evolution of the
RDM is governed by equations containing no memory kernel.  In the original
Redfield theory \cite{redf57,redf65} the secular approximation was 
performed. In this approximation it is assumed that every element of the
RDM in the energy representation is coupled only to those
elements that oscillate at the same frequency. In the present study we do
not perform this additional approximation.  

For larger systems numerical implementations for solving the Redfield
equation are numerically very demanding and therefore require that one
finds the most appropriate way to perform the time evolution of the RDM.
Straightforward one can construct and directly diagonalize the Liouville
superoperator. For processes with a time-independent Hamiltonian the rates,
i.e.\ the characteristic inverse times of an exponential decay of the
occupation probability of the excited states, can be obtained in this way. In
such an approach a huge number of floating point operations will be
involved and the overall computational effort will scale as ${\cal N}^6$ 
where ${\cal N}$
is the size of the basis of vibronic states.  Furthermore the direct
diagonalization can be numerically unstable, but nevertheless has been
successfully used (see e.g.\ \cite{jean92}).  Another strategy suggests
solving ${\cal N}^2$ ordinary differential equations and requires products between
the Liouville superoperator and the RDM which scale as
${\cal N}^4$ (see e.g. \cite{guo99}). This is also numerically demanding for
larger systems.  Assuming a bilinear system-bath coupling the numerical
effort can be reduced considerably by rewriting the Redfield equation in
such a form that only matrix-matrix multiplications are needed
\cite{poll94} rather than applying a superoperator onto the RDM.
Hence, a computational time scaling of ${\cal N}^3$
and a storage requirement of
${\cal N}^2$ is achieved. In the present paper our numerical studies will be based
on this approach.

If one wants to reduce the scaling of the numerical effort with increasing
number of basis functions even more one has to go to stochastic wave
function methods \cite{dali92,garr94,breu95a,breu95}.  They prescribe
certain recipes to unravel the Redfield equation and to substitute the
RDM by a set of wave functions which evolve partially
stochastically in time. The method will have the typical scaling of the
well developed and optimized wave function propagators, i.e. ${\cal N}^2$.  It has
been applied, for example, to electron transfer systems
\cite{wolf95,wolf96,wolf98} and shown to give accurate results.  Direct
solutions and the stochastic wave packet simulations have already been
compared numerically \cite{saal96,breu97}. All these first studies were
restricted to dissipation operators with Lindblad form \cite{lind76}.
Breuer et al. \cite{breu99} showed that the stochastic wave function
approach can also be applied to Redfield master equations without the
secular approximation and for non-Markovian quantum dissipation.  In
particular for complex systems with a large number of levels its practical
application is very advantageous. Between the direct and the stochastic
methods to solve the Redfield equation the accuracy differs especially
because direct RDM integrators are numerically ``exact'' while the
stochastic wave function simulation methods have statistical error. The
scope of this work are small and medium size problems. Therefore we compare
the different numerical algorithms for a direct integration of the Redfield
equation. But when using stochastic methods for density matrix
propagation one has to solve
Schr\"odinger-type equations with a non-Hermitian Hamiltonian. For that
purpose the same algorithms as investigated here can be used. In this sense
the present study is also of importance for solving Redfield equations by
means of stochastic methods.

Here we use different numerical schemes to solve the Liouville-von Neumann
equation. The performance of the well-known Runge-Kutta (RK) scheme is
studied in two different implementations: as given in the {\it Numerical
  Recipes} \cite{pres92} and by the {\it Numerical Algorithms Group}
\cite{nag}.  Compared to these general-purpose solvers are more special
algorithms which have been applied previously to the time evolution of wave
packets and density matrices. For density matrices these are the
Short-Iterative-Arnoldi (SIA) propagator \cite{poll94,poll96}, the
short-time Chebyshev polynomial (CP) propagator \cite{guo99}, and the
Newtonian polynomial (NP) propagator \cite{berm92,ashk95}. The latter
propagator is also used as a reference method because of its high accuracy.
In addition, a symplectic integrator (SI), which was originally developed
for solving classical equations of motion and extended to wave
packet\cite{gray94,gray96} and density matrix propagation \cite{kaly00}, is
tested.

Besides the propagation algorithm one has to determine the appropriate
representation in which to calculate the elements of the RDM and the
Liouville superoperator. The choice depends strongly on the type of
physical problem that one considers.  Coordinate (grid) representation has
an advantage when dealing with complicated potentials, e.g.  non-bonding
potentials. For the problem of dissociation dynamics in the condensed phase
the grid representation has been applied based on a Lindblad-type master
equation \cite{burg98}. Other examples using grids are works of Gao
\cite{gao98} and Berman et al. \cite{berm92}.  Convenient treatment of
electron transfer dynamics is done in state representation, because one can
model the system using a set of harmonic diabatic potentials \cite{kueh94}.
Other authors \cite{poll94,jean95,jean98} choose the adiabatic eigenstates
of the whole system as a basis set to treat similar problems. A comparative
analysis of the benefits and drawbacks of the diabatic and adiabatic
representation in Redfield theory will be given elsewhere \cite{kond:u}.
From the viewpoint of numerical efficiency we focus on both representations
in the present article.

The paper is organized as follows.  In the next section we will make a
short introduction to the model system and a discussion on the versions of
the Redfield equation and the numerical scaling that they exhibit. In
Section \ref{sec:prop-methods} the methods for propagation used in this
work are briefly reviewed. In Section \ref{sec:performance} we compare the
efficiency of several propagators in solving a simple electron transfer
problem with multiple levels. A summary is  given in the last section.

\section{The Redfield equation and its scaling properties}
\label{sec:eom}

In the RDM theory the full system is divided into a relevant system part
and a heat bath. Therefore the total Hamiltonian consists of three terms
-- the system part $H_{\rm S}$, the bath part $H_{\rm B}$ and the
system-bath interaction $H_{\rm SB}$:
\begin{equation}
H = H_{\rm S} + H_{\rm B} + H_{\rm SB}.
\label{eq:Hamiltonian}
\end{equation}

The separation into system and bath allows one to formulate the system
dynamics, described by the Liouville-von-Neumann equation, in terms of the
degrees of freedom of the relevant system. In this way one loses the
mostly unimportant knowledge of the bath dynamics but gains a great
reduction in the size of the problem.  Such a reduction together with a
second order perturbative treatment of the system-bath interaction $H_{\rm
  SB}$ and the Markov approximation leads to the Redfield equation
\cite{redf57,redf65,blum96,may00}:
\begin{equation}
  \dot{\rho} = - \frac{i}{\hbar}[H_{\rm S},\rho] + {\cal R}\rho 
={\cal L} \rho.
\label{eq:redfield}
\end{equation}
In this equation $\rho$ denotes the reduced density operator and ${\cal R}$
the Redfield tensor. If one assumes bilinear system-bath coupling with
system part $K$ and bath part $\Phi$
\begin{equation}
H_{\rm SB} = K \Phi
\label{eq:bath-coupling}
\end{equation}
one can take advantage of the following decomposition \cite{poll94,may00}:
\begin{equation}
\dot{\rho} = - \frac{i}{\hbar}\left[ H_{\rm S},\rho \right]
+ \frac{1}{\hbar^2}
\{
\left[ \Lambda\rho,K \right]+
\left[ K,\rho\Lambda^{\dagger} \right]
\}.
\label{eq:pf-form}
\end{equation}
Here the system part $K$ of the system-bath interaction and $\Lambda$
together hold equivalent information as the Redfield tensor ${\cal R}$.
The $\Lambda$ operator can be  written in the form

\begin{equation}
\Lambda=K\int\limits_0^{\infty}d\tau
\langle\Phi(\tau)\Phi(0)\rangle e^{i\omega\tau}
=K C(\omega)~.
\label{eq:lambda}
\end{equation}

The two-time correlation function of the bath operator
$C(\tau)=\langle\Phi(\tau)\Phi(0)\rangle$ and its Fourier-Laplace image
$C(\omega)$ can be relatively arbitrarily defined and depend on a
microscopic model of the environment.  Different classical and quantum bath
models exist. Here we take a quantum bath \cite{poll96}, i.e. a large
collection of harmonic oscillators in equilibrium, that is characterized by
a Bose-Einstein distribution and a spectral density $J(\omega)$:
\begin{equation}
C(\omega)=2\pi\left[1+\left(e^{\hbar\omega/k_B T}-1 \right)^{-1}\right] 
\left[J(\omega)-J(-\omega)\right]~.
\label{eq:corr}
\end{equation}

\begin{figure}
\mbox{
\psfig{figure=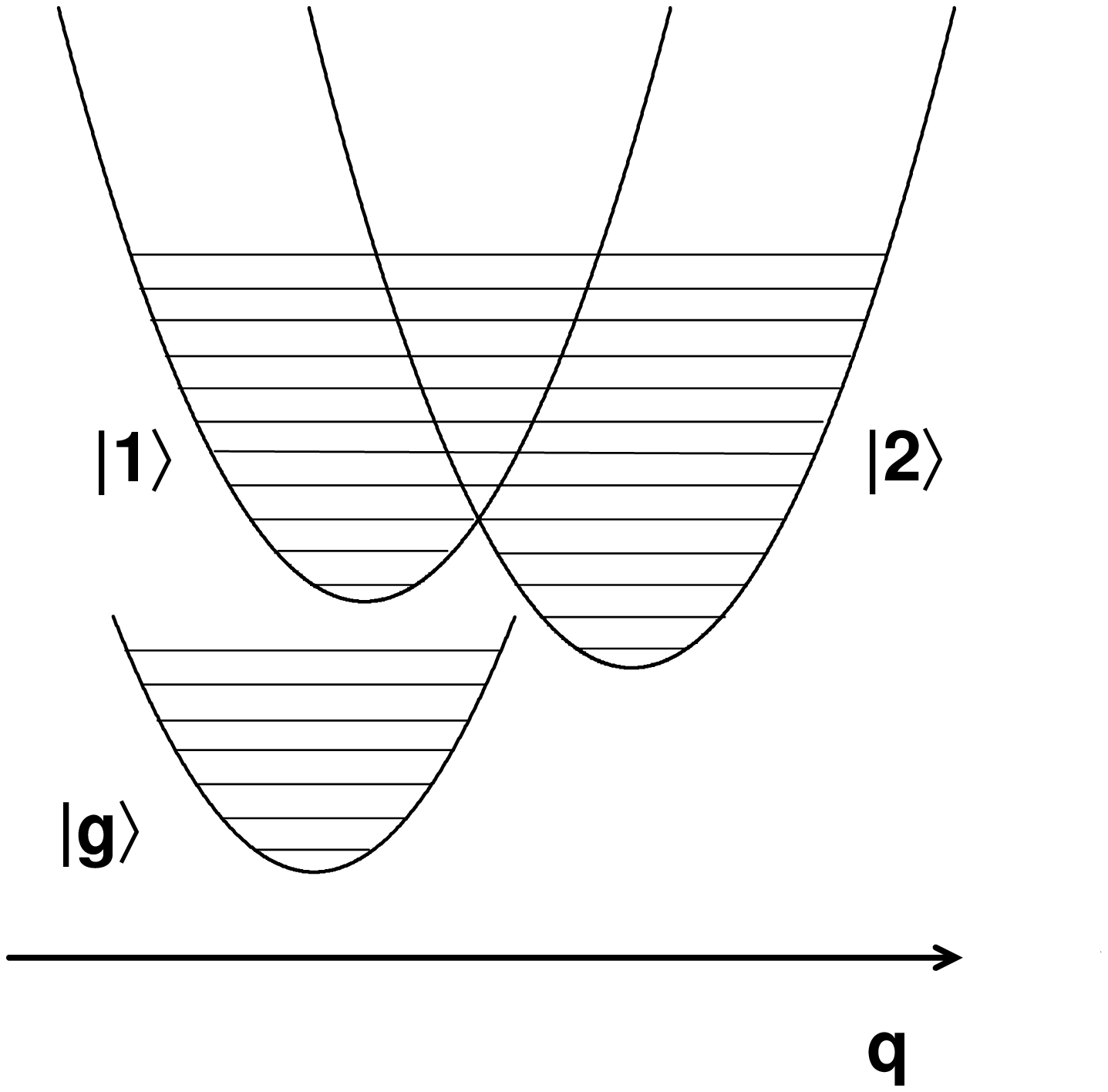,width=7cm}
\psfig{figure=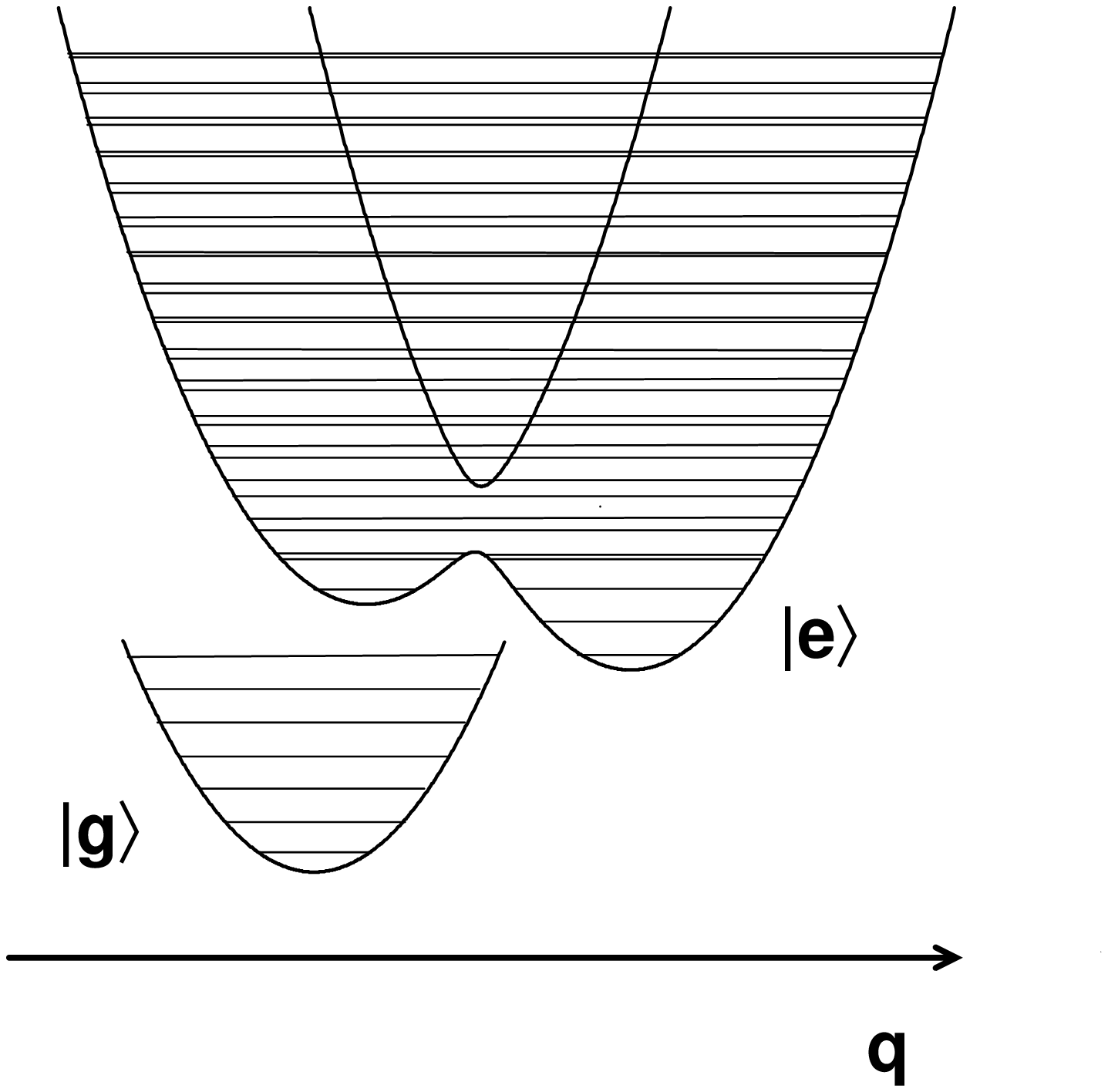,width=7cm}
}
\vspace{1cm}
\caption{Potential energy surfaces (PESs) for a model electron transfer
system. Diabatic PESs are plotted on the left side, and the PES of the
adiabatic excited state $|{\rm e}\rangle$
on the right side.}
\label{fig:PES}
\end{figure}

Now we look for bases that span the degrees of freedom of the relevant
system. Let
us consider atomic or molecular centers $m$ at which the electronic states
$|m\rangle$ of the system are localized.  Their potential energy surfaces
(PESs) will be approximated by harmonic oscillator potentials, displaced
along the reaction coordinate $q$ of the system. As such a coordinate one
can, for example, choose a normal mode of the relevant system part which is
supposed to be strongly coupled to the electronic states. The centers are
coupled to each other with constant coupling $v_{mn}$. For example two such
coupled centers are sketched in Fig.~\ref{fig:PES}. The coupled surfaces
$|1\rangle$ and $|2\rangle$ are assumed to describe excited electronic
states.  The electron transfer takes place after an excitation of the
system from its ground state $|\mbox{g}\rangle$.

Using this microscopic concept we define $K$ as the system's
coordinate operator
\begin{equation}
K = q = \sum\limits_m \left( 2\omega_m {\cal M}/\hbar \right)^{-1/2} 
\left( a_m^{\dagger} +a_m \right) |m \rangle\langle m|~
\label{eq:ka}
\end{equation}
where $a_m$ and $a_m^{\dagger}$ are the boson operators for the normal
modes at the  center $m$, ${\cal M}$ is the reduced mass and $\omega_m$
are the
eigenfrequencies of the oscillators. In the same picture the Hamiltonian of
the relevant system reads
\begin{equation}
H_{\rm S}= \sum\limits_{mn} 
\{ \delta_{mn}
\left[
U_m+
 \left( a_m^{\dagger} a_m + \frac{1}{2} \right) \hbar\omega_m
 \right] 
+ \left(  1-\delta_{mn} \right) v_{mn} \} |m \rangle\langle n|~.
\label{eq:ham}
\end{equation}
From this point on we consider two possible state representations in
order to calculate the matrix elements of $\rho$ and the operators
$H_{\rm S}$,  $K$ and $\Lambda$.
The diabatic (local) basis is a direct product of the eigenstates 
$|M\rangle$ of the harmonic oscillators and the
relevant electronic states $|m\rangle$ (Fig.~\ref{fig:PES}, left panel).
 The intercenter coupling
$v_{mn}$ gives
rise to off-diagonal elements of the Hamiltonian matrix
\begin{equation}
\langle mM|H_{\rm S}|nN\rangle=
\left[U_{m}+\left( M+\frac{1}{2}\right)\hbar\omega_{m}\right]\delta_{mn}\delta_{MN}
+\left( 1-\delta_{mn}\right)v_{mn}\langle mM|nN\rangle~,
\label{eq:dia-ham}
\end{equation}
where $U_m$ are the energies of the minima of the diabatic PESs.  Other
important properties of the diabatic representation are the equidistance in
the level structure and the diagonal form of the system-bath interaction
operator $H_{\rm SB}$. This determines the tridiagonal band form of the
operator $K$.

When neglecting the influence of the intercenter coupling on the
dissipation a very efficient numerical algorithm can be derived
\cite{may92,kueh94}. Of course, for strong intercenter couplings the
populations $\langle mM |\rho | mM \rangle$ at long times deviate from
their expected equilibrium values.  But even for very small couplings there
are cases in which the population does not converge to its equilibrium
value \cite{kond:u}. Therefore this neglect of the influence of the
intercenter coupling on the dissipation has to be handled with care.  On
the other hand this approximation makes the extension of the present
electron transfer model to many modes conceptually much easier
\cite{wolf95,wolf96}.

\begin{figure}
\psfig{figure=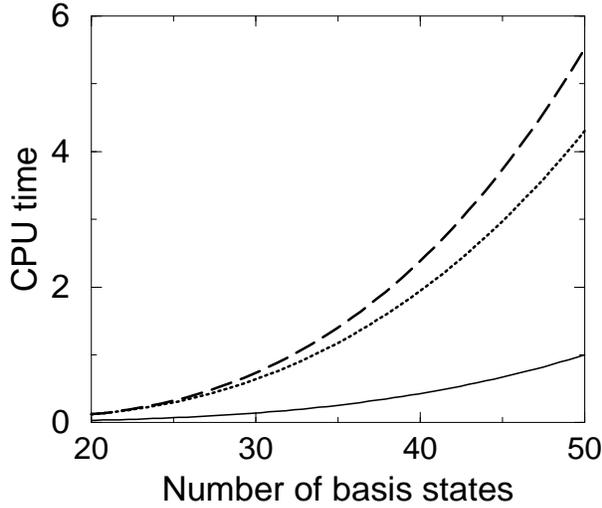,width=8cm}
\caption{Scaling behavior of the product ${\cal L} \rho(t)$. Solid line
  -- tridiagonal form of $K$ in diabatic representation, dotted line --
  adiabatic representation, dashed line -- diabatic representation with
  full matrix-matrix multiplications.  The CPU time is scaled so that it is
  equal to 1 for ${\cal N}=50$ in diabatic representation.}
\label{fig:cpun}
\end{figure}

The matrix elements of the operators in the dissipative part of
Eq.~(\ref{eq:pf-form}) read
\begin{equation}
\langle mM|K|nN\rangle=
\left( 2\omega_m {\cal M}/\hbar \right)^{-1/2}\delta_{mn}
\left( \delta_{M+1,N}\sqrt{M+1}+\delta_{M-1,N}\sqrt{M}\right)
\label{eq:dia-ka}
\end{equation}
and
\begin{equation}
\langle mM|\Lambda|nN\rangle=\langle mM|K|nN\rangle C(\omega_{mMnN})~.
\label{eq:dia-lambda}
\end{equation}
In Eq.~(\ref{eq:dia-lambda}) $\omega_{mMnN}$ denote the transition
frequencies of the system
\begin{equation}
\hbar\omega_{mMnN}=\langle mM|H_{\rm
S}|mM\rangle-\langle nN|H_{\rm S}|nN\rangle.
\label{eq:trans-freq}
\end{equation}
Since the system can emit or absorb only at the eigenfrequencies of the
system oscillators $\omega_m$ the spectral density of the bath $J(\omega)$
is effectively reduced to a few discrete values
$J(\omega)=\sum_m \gamma_m\delta(\omega-\omega_m)$.
The advantage of this approach lies in the scaling behavior with the number
of basis functions. As shown in Fig. \ref{fig:cpun} it scales like ${\cal
N}^{2.3}$
where ${\cal N}$ is the number of basis functions. This is far better than 
the scaling without neglecting the influence of the intercenter coupling
on the dissipation as described below.

If the system Hamiltonian $H_{\rm S}$ is diagonalized it is possible to use
its eigenstates as a basis (Fig.~\ref{fig:PES}, right panel), in which to
calculate the elements of the operators in Eq.~(\ref{eq:pf-form}). Of
course there will be no longer any convenient structure in $K$ or
$\Lambda$, so that the full matrix-matrix multiplications are inevitable.
For this reason the computation of ${\cal L} \rho (t)$ scales as ${\cal N}^3$, 
where ${\cal N}$ is the number of eigenstates
of $H_{\rm S}$.  There appears to be a minimal number ${\cal N}_0$ below
which the diagonalization of $H_{\rm S}$ fails or the completeness relation
for $|mM\rangle$ is violated.  Nevertheless, the benefit of this choice is
the exact treatment of the system-bath interaction.  Denoting the unitary
transformation  that diagonalizes $H_{\rm S}$ by $U$ one has
\begin{equation}
{\bf \epsilon}=U^{\dagger}H_{\rm S}U
\label{eq:u-trans}
\end{equation}
where ${\bf \epsilon}$ is a diagonal matrix containing the eigenvalues. In
this way it is straightforward to obtain the matrices for $\rho$ and $K$.
Equation (\ref{eq:corr}) for $C(\omega)$ still holds but a new definition
of the spectral density $J(\omega)$ is necessary because of the
non-equidistant adiabatic eigenstates. The bath absorbs over a large region
of frequencies and this is characterized in the model by $J(\omega)$.  One
needs the full frequency dependence of $J(\omega)$ which we take to be of
Ohmic form with exponential cut-off:
\begin{equation}
J(\omega)=\eta \Theta(\omega)\omega e^{-\omega/\omega_c}.
\label{eq:ohmic}
\end{equation}
Here $\Theta$ denotes the step function and $\omega_c$ the cut-off
frequency.  In this study all system oscillators have the same frequency
$\omega_1$ (see Table \ref{tab:parameters}) and the cut-off frequency
$\omega_c$ is set to be equal to $\omega_1$. The normalization prefactor
$\eta$ is determined such that
\begin{equation}
\int\limits_0^{\infty}d\omega J(\omega)=\gamma_1~.
\label{eq:normalization}
\end{equation}
Equation~(\ref{eq:ohmic}) together with Eq.~(\ref{eq:corr}) yields the
correlation function in adiabatic representation.

The introduced representations allow us to consider the numerical effort
for a single computation of the right hand side of
Eq.~(\ref{eq:pf-form}), i.e.~of
${\cal L} \rho (t)$.
In the diabatic representation its computation can be approached using two
different algorithms. It is possible to perform matrix-matrix
multiplication only on those elements of $K$ and $\Lambda$ which have
nonzero contributions to the elements of ${\cal L} \rho (t)$
(Fig.~\ref{fig:cpun}, solid line). This is advantageous because of the
tridiagonal form of $K$ in diabatic representation and shows the best
scaling properties, namely ${\cal N}^{2.3}$. In the same representation but
performing the full matrix-matrix multiplications in Eq.~(\ref{eq:pf-form})
(Fig.~\ref{fig:cpun}, dashed line) the scaling behavior is slightly worse
than the same operation in adiabatic representation (Fig.~\ref{fig:cpun},
dotted line). This is due to the non-diagonal Hamiltonian $H_{\rm S}$ in
the former case that makes the computing of the coherent term (see
Eq.~(\ref{eq:redfield})) more expensive.  Below we will take the full
matrix-matrix multiplication to evaluate ${\cal L} \rho (t)$ in both
representations. We do this to concentrate on the various propagation
schemes, not the unequal representations. Nevertheless there are
performance changes in the different representations because of the
disparate basis functions and forms of the operators in these basis
functions.

\section{The different propagation schemes}
\label{sec:prop-methods}
\subsection{Runge-Kutta method}
The RK algorithm is a well-known tool for solving ordinary differential
equations. Thus, this method can be successfully applied to solve a set of
ordinary differential equations for the matrix elements of
Eq.~(\ref{eq:pf-form}). It is based on a few terms of the Taylor
series expansion. In the present work we use the FORTRAN77 implementation
as given in the {\it Numerical Recipes} \cite{pres92} which is a fifth-order
Cash-Karp RK algorithm and will be denoted as RK-NR.  As alternative
the RK subroutine from the {\it Numerical Algorithms Group} \cite{nag}
which is based on RKSUITE \cite{bran:u} was tested with a 4(5) pair.
It will be referred to as RK-NAG.  Both RK-NAG and RK-NR involve terms of
fifth order and use a prespecified tolerance $\tau$ as an input parameter
for the time step control.  The tolerance $\tau$ and the accuracy of the
calculation are not always simply proportional.  Usually decreasing $\tau$
results in longer CPU times.

In our previous work \cite{schr00} a time step control mechanism different
from those used in RK-NAG and RK-NR was tested. Discretizing the time
derivative in Eq.~(\ref{eq:redfield}) and requiring
\begin{equation}
  \left| \frac{\rho(t_{i+1})-\rho(t_i)}{\Delta t} +
    {\cal L} \rho(t_i) \right| < \tau
\label{eq:rk2}
\end{equation}
one only has to call the propagation subroutine once and to store the
previous RDM.  In addition one has to calculate the action of the Liouville
superoperator ${\cal L}$ onto the RDM but the numerical effort for this is
small compared to a call of the propagation subroutine. It was shown in
Ref.~\cite{schr00} that this time step control is the most efficient for
propagation with the coherent terms in Eq.~(\ref{eq:redfield}) only but
disadvantageous for problems with dissipation. This is the reason why we do
not include this algorithm in the present study.

\subsection{Short Iterative Arnoldi propagator}
The SIA propagator \cite{poll94,poll96} is a generalized version of the
Short Iterative Lanczos propagator \cite{lefo91} to non-Hermitian
operators. With the Short Iterative Lanczos algorithm the wave function can
be propagated by approximating the time evolution operator in Krylov space,
which  is generated by  consecutive
multiplications of the Hamiltonian on the wave function. In analogy the
Krylov space within the SIA method is constructed by recursive applications
of the Liouville superoperator onto the RDM $\rho_n ={\cal L}^n
\rho(t)$.  In this way it is tailored for the RDM at every moment in
time.  The Liouville superoperator, denoted by $l$ in Krylov space, has
Hessenberg form
\begin{equation}
{\bf {\cal L}} \approx {V} l V^{\rm T}~,
\label{eq:sia1}
\end{equation}
where the orthogonal transformation matrix $V$ is constructed 
iteratively using the so-called Lanczos procedure \cite{poll94}. 
The Krylov representation $l$ can be easily 
diagonalized to $L$ with the help of a transformation matrix
$S$:
\begin{equation}
e^{{\cal L}t}
\approx
V S e^{L t} S^{-1} V^{\rm T}~.
\label{eq:sia2}
\end{equation}
Since the diagonalization is performed in the Krylov space the numerical
effort depends on its dimension which can be chosen small in practice.
Having thus derived a diagonal operator $e^{{ L}t}$ the calculation of
$\rho(t)$ is straightforward.

\subsection{Symplectic integrator}

The SIs were originally developed for solving classical equations of motion
\cite{ruth83}. The time evolution of a classical Hamiltonian system can be
viewed as a canonical transformation and SIs are sequences of canonical
transformations. Recently it was shown that the time evolution of wave
packets \cite{gray94,gray96} and  density matrices \cite{kaly00} can also be
performed using SIs. In order to rewrite the Redfield equations in the form
of coupled canonical variables that are analogous to classical equations
of motion one defines the functions\cite{kaly00}
\begin{eqnarray}
Q(t)&=&\rho(t)~, \\
P(t)&=&  \dot{\rho}(t)~,
\end{eqnarray}
the operator 
\begin{eqnarray}
  \label{w}
W&=& -\frac1{\hbar^2}{\cal L}^2  ~,
\end{eqnarray}
and the Hamiltonian function
\begin{eqnarray}
  \label{hf}
  G(Q,P)=\frac12[P^TP+Q^TWQ]~.
\end{eqnarray}
Doing so one obtains equations of motion analogous to the classical ones
\begin{eqnarray}
  \label{cem}
  \frac{d}{dt}P(t)&=&-\frac{\partial G(Q,P)}{\partial Q}=- W Q(t)~, \\
  \frac{d}{dt}Q(t)&=&\frac{\partial G(Q,P)}{\partial P}= P(t)~. 
\end{eqnarray}
Rewriting this into the SI algorithm of order $m$  yields \cite{kaly00}
\begin{eqnarray}
  \label{SIalgo}
&&P_i=P_{i-1}+\frac{b_i \Delta t}{\hbar^3}{\cal L}^2 Q_{i-1}\\
&&Q_i=Q_{i-1}+\frac{a_i \Delta t}{\hbar} P_i 
\end{eqnarray}
for $i=1, \ldots ,m$.  Different sets of coefficients $\{a_i\}$ and
$\{b_i\}$ are given in the literature. Here we choose the McLachlan-Atela
fourth-order method \cite{mcla90}. The coefficients for this method are
listed in Ref.~\cite{gray94b}. A comparison of the McLachlan-Atela
fourth-order method with the McLachlan-Atela third-order method
\cite{mcla90} and Ruth's third-order method \cite{ruth83} has been given
elsewhere \cite{kaly00}.

\subsection{Newton polynomial scheme}
Another way to solve Eq.\ (\ref{eq:redfield}) is by a polynomial expansion
of the time-evolution operator. Such methods are well established and
approved for wave-function propagation \cite{lefo91,ashk95}.  Recently the
Faber \cite{huis99} and NP \cite{berm92} algorithms have been applied to
propagate density matrices and it has been shown that they behave very
similarly \cite{huis99}.  The main idea of the NP method is the
representation of the Liouville superoperator by a polynomial
interpolation
\begin{equation}
e^{{\cal L} t} 
\approx {\cal P}_{N_p-1}({\cal L}) 
\equiv \sum\limits_{n=0}^{N_p-1} a_n \rho_n
= \sum\limits_{n=0}^{N_p-1} a_n \prod\limits_{j=0}^{n-1}({\cal 
L}-\lambda_j)
\label{eq:newt}
\end{equation}
of order $N_p$ where the $\rho_n$ are computed recursively and $a_n$ are the $n$-th divided
differences. The interpolation points $\lambda_j$ can be chosen to form a
rectangular area in the complex plane (see Fig.~\ref{fig:eigenvalues})
which contains all eigenvalues of ${\cal L}$.  This interpolation scheme is
uniform, i.e., the accuracy in energy space is approximately the same in the
whole spectral range of ${\cal L}$. This is in contrast to schemes such as
the SIA propagator which are nonuniform approximations.  A consequence of
this property is the very high accuracy which can be achieved with uniform
propagators.  This is why we take a high-order NP expansion as reference
solution.  Since the quality of the approximation of the time evolution
operator is equivalent to a scalar function with the same interpolation
points $\lambda_j$, one can, before performing the actual calculation,
check the accuracy on a scalar function. For the calculation with the NP
propagator we set the truncation limit of the expansion to $10^{-15}$,
i.e.,\ the sum in Eq.~(\ref{eq:newt}) is truncated when the residuum
fulfills $a_n||\rho_n|| < 10^{-15}$ \cite{ashk95}.

\subsection{Chebyshev polynomial scheme}
\begin{figure}
\psfig{figure=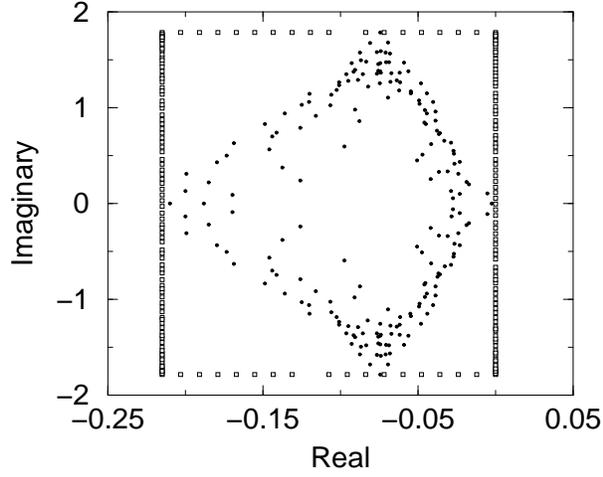,width=8cm}
\caption{Scaled spectrum $L/L^-$ of the Liouville superoperator for the model
  of electron transfer.  Approximate eigenvalues obtained in Krylov
  subspace are plotted as dots.  Open squares denote the
  interpolation points $\lambda_j$ for the NP scheme.}
\label{fig:eigenvalues}
\end{figure}

As a last contribution to the present study we will examine the CP
propagator.  Recently it was studied by Guo et al. \cite{guo99} for density
matrices.  The Liouville superoperator is approximated by a series of CPs
$T_k(x)$.  Generally the CPs diverge for non-real arguments. For propagators
of the kind $e^{-i H t}$ it has been shown \cite{ashk95} that the CPs may
tolerate some imaginary part in the eigenvalues of $H$. The stability
region has the form of an ellipse with a center at the origin and a very
small half-axis in imaginary directions \cite{ashk95}. In contrast, the
eigenvalues of the Liouville superoperator are spread over the negative
real half of the complex plane and symmetrically with respect to the real
axis (see Fig.~\ref{fig:eigenvalues}). The real components for the system
that we consider are one order of magnitude smaller than the imaginary
components. This is why we make the expansion along the imaginary axis and
use an expression similar to that already applied to wave function
propagation \cite{lefo91}:
\begin{equation}
e^{{\cal L} t}
\approx 
e^{L^+\Delta t}\sum\limits_{n=0}^{N_p-1} (2-\delta_{n0})J_n(L^-\Delta t)
T_n({\tilde L})~.
\label{eq:cheb}
\end{equation}
Here the expansion coefficients $J_n$ are the Bessel functions of
the first kind, and ${\tilde L}$ is the appropriately scaled Liouville 
superoperator: ${\tilde L}=({\cal L}-L^+)/L^-$, where $L^-$ and $L^+$
are the half span and the middle point of the spectrum of ${\cal L}$.
Since the spectrum is symmetric with respect to the real axis, $L^+=0$. The time evolution of $\rho$ is given by
\begin{equation}
\rho(t+\Delta t)
\approx \sum\limits_{n=0}^{N_p-1} (2-\delta_{n0})J_n(L^-\Delta t)
{\tilde \rho}_n~.
\label{eq:cheb-prop}
\end{equation}
The Chebyshev vectors ${\tilde \rho}_n$ are generated by
means of a recurrence procedure:
\begin{equation}
{\tilde \rho}_n = 2{\tilde L}{\tilde \rho}_{n-1}+{\tilde
\rho}_{n-2}, \mbox{~~} {\tilde
\rho}_{0}=\rho(t) \mbox{~and~} {\tilde \rho}_{1}={\tilde L}{\tilde \rho}_0~.
\label{eq:cheb-recur}
\end{equation}

For the CP and NP methods one has to adjust the values of the spectral
parameters $L^-$ and $L^+$. One can obtain some knowledge about the
spectrum of ${\cal L}$  by an approximate diagonalization,
e.g. by Krylov subspace methods. For instance, Fig.~\ref{fig:eigenvalues}
shows an approximate spectrum of ${\cal L}$ appropriately scaled so
that all eigenvalues lie within the rectangle formed by the Newtonian
interpolation points.

\section{Performance of propagation methods}
\label{sec:performance}

The aim of this section is to compare the different numerical methods
described above for propagating the RDM in time. The calculations were
performed for both RK methods with different tolerance parameters $\tau$
and for the SI as well as the NP, CP, SIA propagators with different
timesteps. The number of expansion terms $N_p$ in NP and CP propagators is
170 and 64, respectively. The dimension of the Krylov space for the SIA
method was set to 12 because smaller as well as larger values are less
efficient for the example studied here. All computations were made on
Pentium~III~550 MHz personal computers with intensive use of BLAS and
LAPACK libraries.  The code was compiled using the PGF90 Fortran compiler
\cite{pgf90}.  For estimation of the computational error of all methods the
NP algorithm with 210 terms was chosen as a benchmark. 

In this work we
consider only two centers $m=1,2$ which is the minimal model to describe
the main physics of an electron transfer reaction. A basis size of 16
levels per center satisfies the completeness relation and presents no
difficulties during the diagonalization of $H_{\rm S}$.  The electronic
coupling was $v_{12}=0.1$ eV.  We choose $\gamma_1=\gamma_2=1.57863\times{}{\rm
  10^{-2}~eV}{\rm \AA{}^{-2}}$ and  ${\cal M}=20 m_p$ where $m_p$ is the
proton mass. The temperature $T=298$ K is used. In Table
\ref{tab:parameters} the parameters for the system oscillators are given.

The process that is simulated involves the following scenario. A Gaussian
wave packet is prepared as initial state by a vertical transition from the
lowest vibrational level of the ground electronic state $|{\rm g}\rangle$
to the first (upper) center $|1\rangle$:
\begin{equation}
\rho_{1M1N}(t=0)=
\langle 1M|{\rm g}0 \rangle\langle {\rm g}0|1N \rangle~.
\label{eq:initstate}
\end{equation}
The 
energy distribution of the occupied eigenstates by the wave packet
depends on the displacement $q^0_{\rm g}-q^0_1$
between the PESs of $|{\rm g}\rangle$ and $|1\rangle$.  During the pulse
the two excited electronic states $|1\rangle$ and $|2\rangle$ are assumed
to be decoupled. In this way one can simulate the absorption of
electromagnetic radiation from a pulse with vanishing width.  Right after
the pulse is over, the wave packet starts moving on the excited PESs and
spreading.  The relevant system part begins losing energy to the bath and
dephasing.  The population on the upper center starts decaying. When the
damping is not too strong, as for the model parameter studied here, a
damped oscillation of the population between the two excited PESs can be
seen.  We assume no coupling to the ground state after the pulse.  After a
certain time the system reaches its equilibrium state.

In all cases the RDM was propagated for a total time period of $3\times{}10^{5}$
a.u.\ which is sufficient for complete relaxation to equilibrium.
It was compared to the RDM $\rho_{\rm ref}$ evaluated
by the NP algorithm at the same points in time.  The relative error
$\varepsilon(t)$ of each method at a certain moment in time $t$ has been
estimated using a formula similar to that proposed for wave functions by
Leforestier et al.\cite{lefo91}:
\begin{equation}
\varepsilon (t) = \left| 1 - \frac{{\rm Tr}\left(\rho(t) 
\rho_{\rm ref}(t)\right)}{{\rm Tr}\left(\rho^2_{\rm ref}(t)\right)}
\right|~.
\label{eq:error}
\end{equation}
As the error $\varepsilon$ we define the maximum value of $\varepsilon(t)$
over the total propagation time.  For more details we refer to our previous
paper \cite{schr00}.  Other error measures (see for example
\cite{nett:u,guo99}) can be used as well but they will have the same
qualitative behavior.

\begin{figure}
\psfig{figure=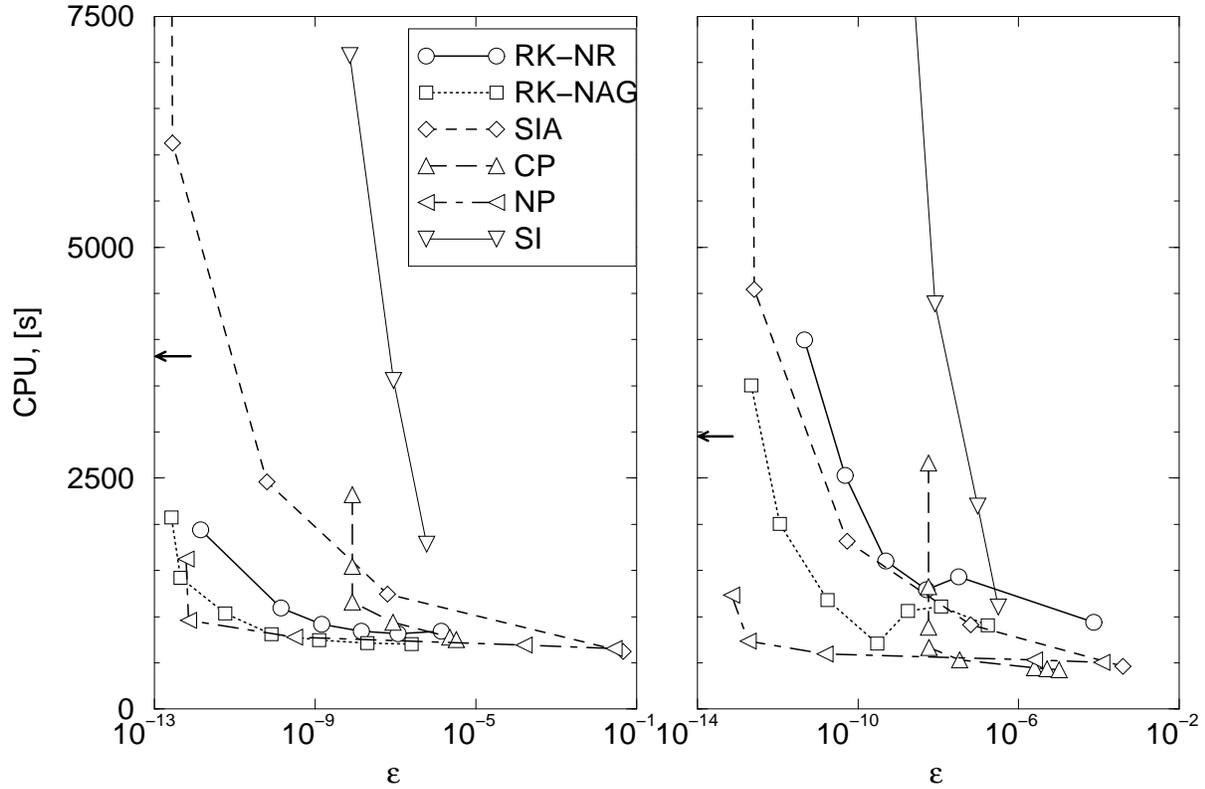,width=16cm}
\caption{Numerical performance of different numerical propagators.
  Results obtained in the diabatic (adiabatic) representation are shown on
  the left (right) plot. The arrows represent the numerical performance for
  the NP propagator with 50 terms and timestep 100 a.u.
}
\label{fig:performa}
\end{figure}

As an index for the numerical effort two possibilities were explored. The
first one is a direct measurement of the CPU time of the total
propagation (Fig.~\ref{fig:performa}). It may look quite different on
other computer architectures or even on the same architecture but under changed
operation conditions. An evidence of the performance (Fig.~\ref{fig:performa})
 will be expressed by means of CPU time versus
 the error $\varepsilon$.
 
\begin{figure}
\psfig{figure=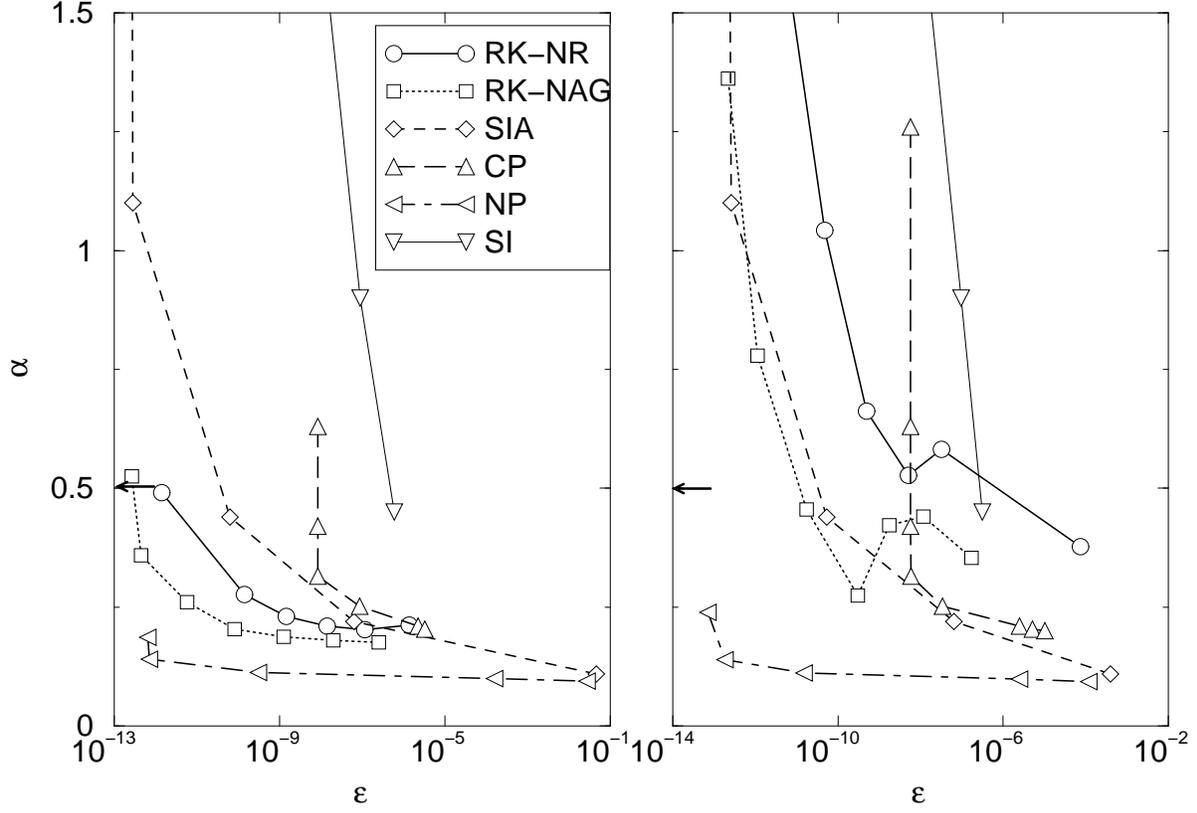,width=16cm}
\caption{Numerical performance of different numerical propagators.
  The numerical effort $\alpha$ is defined in Section
  \ref{sec:performance}. Results obtained in the diabatic (adiabatic)
  representation are shown on the left (right) panel. The arrows represent
  the numerical performance for the NP propagator with 50 terms and
  timestep 100 a.u.  }
\label{fig:performb}
\end{figure}

Another approach to describe the numerical effort
 has been proposed
 \cite{guo99} and  is called efficiency factor. It is defined as
the ratio between the timestep $\Delta t$ and the number of operations
${\cal L} \rho(t)$ within this timestep. Because of the definition it is
a machine independent quantity.
The larger the efficiency factor, the better the performance of the
algorithm. Because the RK algorithms propagate 
with variable timestep we cannot directly use the definition of the
efficiency factor. Instead we define a quantity $\alpha$
as the total number $N_c$ of ${\cal L}\rho(t)$-evaluations 
divided by the total time for the propagation:
\begin{equation}
\alpha=N_c/(N_s\overline {\Delta t})~.
\label{eq:alpha}
\end{equation}
Here $N_s$ denotes the total number of timesteps. 
The inverse of $\alpha$ will have the
meaning of an efficiency factor for an averaged timestep 
$\overline{\Delta t}$.
 We should point out 
that $N_c$
 does not take into account the effort for summation
of the different contributions. In particular in the case of the NP method
the summation of the different terms in the polynomial expansion
Eq.~(\ref{eq:newt}) can be non-negligible. This can be seen in
the different relative performance of the propagators shown in 
Figs.~\ref{fig:performa} and
\ref{fig:performb}.
We consider both the CPU time and the quantity
$\alpha$ as measures of the numerical effort.

Contributions from the algorithm to calculate ${\cal L} \rho(t)$ also
influence the CPU time. As discussed above, in all computations represented
in Figs.~\ref{fig:performa} and~\ref{fig:performb} the full matrix-matrix
multiplications in Eq.~(\ref{eq:pf-form}) were performed. The performance of
the CP, NP, SI and SIA methods is only influenced very little by the choice
between diabatic and adiabatic representation.  Both RK implementations are
less efficient in the adiabatic than in the diabatic representation, though
the RK-NAG scheme has still the best performance besides the NP algorithm.
The RK-NR scheme has an advantage for computation in diabatic rather than
in adiabatic representation especially for medium precision requirements.
In that range the performance curves of the RK methods exhibit a shoulder
for the adiabatic case which seems to result from a numerical artifact.

Because the error of the SIA algorithm is not uniformly distributed in
energy space \cite{kosl94} we could expect some difference in its
performance in diabatic and adiabatic representation.  But because the
coupling $v_{12}$ chosen here is not very large, the eigenstates of the
coupled system are just slightly disordered (see Fig.~\ref{fig:PES}, right
plot) and hence the performance of the SIA algorithm is almost not changed.

The uniformity, stability and high accuracy of the CP propagator for wave
functions is well known \cite{lefo91,kosl94,nett:u}. The CP approach to
density matrix propagation was introduced by Guo and Chen \cite{guo99}.
Using a damped harmonic oscillator as model system and starting from a pure
state they established that the relative error can reach the machine
precision limits ($10^{-15}$) for sufficiently small stepsize. However, for
the system of coupled harmonic potentials studied here and using an initial
RDM with non-zero off-diagonal elements the error saturates at
$\varepsilon\approx 10^{-8}$ (see Fig.~\ref{fig:performa}). It was not
possible to decrease this saturation limit of $\varepsilon$ neither by
increasing the order of the CP nor by decreasing the timestep. This
saturation limit seems to depend strongly on the imaginary part of the
eigenvalues of ${\cal L}$.  For large timesteps the CP method loses its
stability and one needs to estimate the efficiency range of $N_p$, $\Delta t$
and $L^-$.  Turning off the dissipation we could reach much higher accuracy
with the CP propagator as expected.

The SI is easy to implement. The expansion coefficients are fixed and can
be taken from literature. At the same time the fixed coefficients seem to
limit the accuracy. For not too high accuracy the performance of the SI is
as good as that of the other propagators in adiabatic representation. In
diabatic representation its performance is a little worse. But we were not
able to achieve very high accuracy with this method.  This might be due to
the special version, the fourth order McLachlan-Atela method, which we
chose.

As already highlighted \cite{huis99} the NP scheme is very stable for
arbitrary spectral properties of ${\cal L}$. The only restriction is that
the spectrum must be confined within the area formed by the interpolation
points. In our investigation the NP propagator performs with a good
accuracy for timesteps of 1500 a.u. ($N_p=170$) which is 10 times larger
than the step size of the CP scheme.  Higher order expansions might be even more
efficient but the numerical implementation gets tricky and easily unstable.
For timesteps of 100 a.u.  and $N_p=50$ the NP algorithm is already
numerically exact but computational very expensive (see the arrows in
Fig.~\ref{fig:performa} and Fig.~\ref{fig:performb}).  For problems with
time-dependent Hamiltonians (e.g. non-stationary external fields with
relatively small amplitude) the RK and SIA methods will be more efficient
with small timestep.

At the end we should point out that there exists no ultimate method to
determine the performance of a certain numerical approach which could be
valid for different platforms.  Tuning and optimization features are
generally not portable and this may cause even different scaling behavior
and hence a different method of preference. That is the reason why the
generality of the results is limited to similar computation platforms and
even to systems with similar properties of the corresponding Liouville
superoperator.  But on the other hand this study can give hints on the
performance of the different algorithms in general.
 
\section{Summary}
\label{sec:conclusion}
In the present work an estimation of the numerical efficiency of several
methods for density matrix propagation has been given. The example of
electron transfer in a two-center system has been used for this purpose. A
specific measure of the numerical effort has been introduced in order to
compare methods with fixed timestep and such ones with timestep control
(RK). Besides the method of reference (NP) the RK-NAG approach shows best
performance for both cases of adiabatic and diabatic representation. The
advantage of the SIA propagator is that the accuracy improves with
decreasing the timestep in all cases we investigated. That is not the case
with the CP propagator which exhibits a saturation of accuracy and is
therefore not convenient for very small timesteps. The easy-to-implement SI
gives reasonable performance for not too high accuracy. The present SI
seems to be limited in accuracy due to the fixed coefficients.

 The present studies were restricted to
state bases.  Of course similar calculations can be done on a grid which is
especially useful for complicated or unbound potentials. In these cases
another propagator, the split operator \cite{garr95}, should be taken into
account.  This operator has the advantage that its performance does not
(directly) depend on the spectral range of the Hamiltonian or Liouville
operator. So it may perform very well for problems with a large spectral
range although it cannot be applied to operators which have mixed terms in
coordinate and momentum operators. The use of the mapped Fourier
method \cite{fatt96} may reduce the number of grid points significantly and
first wave packet propagations with this method have been done
\cite{klei99b,nest99}.  Recently the multi-configuration time-dependent
Hartree method has been established to treat density matrix operators
\cite{raab00}. This method might be favorable for multi-dimensional
systems.

The presented methods can be used for various applications in the field of
dissipative molecular dynamics in condensed phases, where the RDM approach
provides a good way of describing processes in systems with one or more
degrees of freedom.  This includes the electron transfer processes
mentioned in the introduction as well as exciton transfer processes
\cite{reng97}.  It can also be used to simulate pump-probe experiments
\cite{wolf98}, surface scattering of atoms \cite{nest99}, etc. Also
coherent control schemes in dissipative environments can be studied
\cite{bard97}.  So the numerical studies given here can be applied to a broad
range of problems in physics, chemistry, and biology.

\acknowledgments We thank C. Kalyanaraman and D. G. Evans for help with the
implementation of the symplectic integrator.  Financial support of the DFG
is gratefully acknowledged.



\begin{table}
\caption{Parameters of the system oscillators used for the computations.}
\begin{tabular}{lrll}
Center $|m\rangle$&  $U_m$, eV& $q^0_m$, ${\rm \AA{}}$&$\omega_m$, eV \\ \hline

$|0\rangle\equiv|g\rangle $ & $-0.60$ & 0.000 & 0.1 \\
$|1\rangle$ & $0.25$ & 0.125 & 0.1 \\
$|2\rangle$ & $0.05$ & 0.363 & 0.1
\end{tabular}
\label{tab:parameters}
\end{table}


\end{document}